\documentclass[11pt,preprint]{aastex}
\usepackage{epsfig}

\newcommand       \Angstrom     {\,{\rm \AA}}          

\newcommand       \cm           {\,{\rm cm}}

\newcommand	  \g		{\,{\rm g}}

\newcommand       \simlt        {\lesssim}
\newcommand       \simgt        {\gtrsim}

\newcommand       \mum          {\,{\rm \mu m}}

\newcommand	  \mre		{m^{\prime}}
\newcommand	  \mim		{m^{\prime\prime}}
\newcommand	  \im		{{\rm Im}}
\newcommand	  \kabs		{\kappa_{\rm abs}}
\newcommand	  \bT		{{\bf T}}
\newcommand	  \cabs		{C_{\rm abs}}
\newcommand	  \cpol		{C_{\rm pol}}
\newcommand	  \cpar		{C_{\rm abs}^{\|}}
\newcommand	  \cper		{C_{\rm abs}^{\bot}} 
\newcommand	  \copol	{C_{\rm pol}^{\rm obl}}
\newcommand	  \cppol	{C_{\rm pol}^{\rm pro}} 
\newcommand	  \ac		{a_{\rm c}}
\newcommand	  \bc		{b_{\rm c}}
\newcommand	  \ecc		{e_{\rm c}}
\newcommand	  \fsil		{f_{\rm sil}}
\newcommand	  \am		{a_{\rm m}}
\newcommand	  \bm		{b_{\rm m}}
\newcommand	  \ecm		{e_{\rm m}}
\newcommand	  \for		{f_{\rm carb}}
\newcommand	  \tauor	{\tau_{3.4}}
\newcommand	  \tausil	{\tau_{9.7}}
\newcommand	  \vsil	        {V_{\rm sil}}
\newcommand	  \vor	        {V_{\rm carb}}
\newcommand	  \Asil	        {A_{\rm 9.7\mu m}}
\newcommand	  \Ach	        {A_{\rm 3.4\mu m}}
\newcommand	  \Psil	        {P_{\rm 9.7\mu m}}
\newcommand	  \Pch	        {P_{\rm 3.4\mu m}}

\newcommand{\figwidth}{4.0in}

\countdef\decade=200
\advance\decade by \year
\countdef\hours=201
\advance\hours by \time
\divide\hours by 60
\countdef\mins=202
\advance\mins by \hours
\multiply\mins by 60
\multiply\hours by 100
\countdef\miltime=203
\advance\miltime by \hours
\advance\miltime by \time
\advance\miltime by -\mins

\shorttitle{The Interstellar 3.4$\mum$ Polarization Feature}

\begin{document}

\title{
 \vspace*{-2.0em}
  {\normalsize\rm submitted to
  {\it The Astrophysical Journal Letters}}\\
  \vspace*{1.0em}
Mid-Infrared Spectropolarimetric Constraints
on the Core-Mantle Interstellar Dust Model
	 }

\author{Aigen Li}
\affil{Theoretical Astrophysics Program,
       The University of Arizona, Tucson, AZ 85719;\\
       and Princeton University Observatory, 
       Peyton Hall, Princeton, NJ 08544;
        {\sf agli@lpl.arizona.edu}}
\and

\author{J. Mayo Greenberg}
\affil{The Raymond and Beverly Sackler Laboratory for Astrophysics, 
Sterrewacht Leiden,\\ Postbus 9513, 2300 RA Leiden, The Netherlands}

\begin{abstract}
In the framework of the silicate core-carbonaceous organic 
mantle interstellar dust model, the bulk of the visual/near-IR 
extinction and the entire polarization are from nonspherical and aligned
core-mantle grains. The 3.4$\mum$ C-H and 9.7$\mum$ Si-O absorption 
features, respectively arising from the hydrocarbon mantle and 
the amorphous silicate core, are expected to be polarized
to a modestly different degree. Spectropolarimetric observations
toward the same lines of sight both in the 3.4$\mum$ region
and in the 9.7$\mum$ region would be of great value to 
test the core-mantle dust model. The fact that the 3.4$\mum$ feature 
is not polarized along the line of sight toward 
the Galactic center source IRS 7 is not yet sufficient
to reject the core-mantle model due to the lack of
spectropolarimetric observation of this region in the
9.7$\mum$ region.  
\end{abstract}

\keywords{dust, extinction --- Galaxy: center --- 
infrared: ISM: lines and bands --- polarization}

\section{Introduction\label{sec:intro}}
Although its total mass is only $\sim 0.1\%$ of
that of galaxies, interstellar dust plays an important role 
in the galactic evolution, the formation of stars and
planetary systems, and possibly, the origins of life.
The physical and chemical nature of interstellar dust is
characterized by its composition, size, shape, and structural form. 
Current major observational information includes the wavelength 
dependent extinction, scattering (albedo), and polarization curves,
the absorption and emission spectra, and the elemental depletions.

The extinction curve is a sharp discriminator of (dominant) size
($\simgt 100\Angstrom$),\footnote{%
  Smaller grains are in the Rayleigh limit at the ultraviolet (UV), 
  and their extinction cross sections per unit volume are 
  independent of size. Infrared (IR) emission provides a stronger
  constraint on their sizes (see Li \& Draine 2001).
  }
but a very poor discriminator of composition.
Interstellar polarization is not a separate subject but should be 
treated as part of the extinction data, except that it indicates that 
some fraction of the interstellar grains are both nonspherical 
and significantly aligned.
It is the extinction (absorption) and emission spectral 
lines instead of the overall shape of the extinction curve 
provides the most diagnostic information on the 
dust composition. Based on the 2175$\Angstrom$ extinction hump
and the ubiquitous 3.4$\mum$ C-H stretching mode, 
9.7$\mum$ Si-O stretching mode, 
and 18$\mum$ O-Si-O bending mode absorption features 
seen in interstellar regions as well as the elemental depletions, 
it has now been generally accepted that interstellar grains 
consist of amorphous silicates and some form of carbonaceous 
materials (see Li \& Greenberg 2002 for a recent review).  

However, the spectroscopic absorption/emission profiles are
unable to tell what structural forms dust grains may take;
i.e., no definite conclusion can be drawn yet on the physical 
relationship between the silicate and carbonaceous grain components.
As a matter of fact, three distinctly different grain morphologies 
are proposed for the three most common dust models --
(1) in the silicate-graphite model (and its updated versions)
the bare silicate and graphite grains are assumed to be physically
separated (Mathis, Rumpl, \& Nordsieck 1977 [hereafter MRN]; 
Draine \& Lee 1984; Siebenmorgen \& Kr\"{u}gel 1992; 
Dwek et al.\ 1997; Li \& Draine 2001; Weingartner \& Draine 2001);
(2) in the silicate core-carbonaceous mantle model the silicate 
grains are assumed to be coated with a layer of carbonaceous
materials in the form of organic refractory (Greenberg 1989a;
D\'{e}sert, Boulanger, \& Puget 1990; Li \& Greenberg 1997) 
or hydrogenated amorphous carbon (HAC; Jones, Duley, \& Williams 1990);
(3) in the composite dust model interstellar grains are taken to be 
fluffy aggregates of small silicates, vacuum, and carbon of various 
kinds (amorphous carbon, HAC, and organic refractories;
Mathis \& Whiffen 1989, Mathis 1996). Assuming reasonable elemental
depletions, all dust models are reasonably successful in reproducing 
the observed interstellar extinction and polarization
curves (MRN; Draine \& Lee 1984; Mathis 1986; Mathis \& Whiffen 1989;
Mathis 1996; Li \& Greenberg 1997; Weingartner \& Draine 2001),
the silicate absorption/polarization features (Draine \& Lee 1984;
Greenberg \& Li 1996; Mathis 1998), and the dust thermal emission 
from the near-IR to submillimeter (D\'{e}sert et al.\ 1990;
Dwek et al.\ 1997; Li \& Draine 2001) except that the composite grains
may be too cold and produce too flat a far-IR emissivity to be
consistent with the observational data (Draine 1994; Dwek 1997).

Very recently, Adamson et al.\ (1999) suggest that spectropolarimetric
studies of the 3.4$\mum$ hydrocarbon and 9.7$\mum$ silicate absorption 
features would provide a powerful constraint on the dust morphology. 
The ubiquitous 9.7$\mum$ silicate absorption feature is often 
found to be polarized; some sources also have the 18$\mum$ 
O-Si-O polarization feature detected (see Smith et al.\ 2000 for 
a summary). Another ubiquitous strong 
absorption band in the diffuse interstellar medium (ISM)
is the 3.4$\mum$ feature, commonly attributed to 
the C-H stretching mode in saturated aliphatic hydrocarbons
(see Pendleton \& Allamandola 2002, Li \& Greenberg 2002 for summaries).
If, for the very same region, only the 3.4$\mum$ feature 
or the 9.7$\mum$ feature is polarized while the other is not,
one can then conclude that the silicate and carbonaceous dust
components are physically separated.
An attempt to measure the polarization of the 3.4$\mum$ 
absorption feature was recently made by Adamson et al.\ (1999) 
toward the Galactic center source IRS 7. 
They found that this feature was essentially unpolarized.  
This appeared to pose a severe challenge to the core-mantle dust model
(see \S\ref{sec:discussion} for discussions).

In this {\it Letter}, we explore the IR absorption and polarization
properties of the silicate core-carbonaceous mantle model with 
emphasis on the polarization of the 3.4$\mum$ C-H feature.
In \S\ref{sec:opct} we discuss the optical properties of 
interstellar hydrocarbon dust material. 
In \S\ref{sec:p2a} we predict the relative degrees of polarization 
of the 3.4$\mum$ C-H and 9.7$\mum$ Si-O features. This will
allow comparison with future observations and provide a potentially
powerful test on the core-mantle dust model. 
In \S\ref{sec:discussion} we compare our model results with 
the available observational data and discuss its implication.
In \S\ref{sec:sum} we summarize our main conclusions.   

\section{Optical Properties of Interstellar Hydrocarbon 
Material\label{sec:opct}}
The interstellar organic hydrocarbon dust component 
reveals its presence in the diffuse ISM through the 3.4$\mum$
absorption feature. Since its first detection in
the Galactic center sources,
it has now been widely seen in the Milky Way Galaxy and other 
galaxies (see Pendleton \& Allamandola 2002 for a summary). 
Although it is generally accepted that this feature is due to 
the C-H stretching mode in saturated aliphatic hydrocarbons, 
the exact nature of this hydrocarbon material remains uncertain. 
Nearly two dozen different candidates have been proposed over 
the past 20 years (see Pendleton \& Allamandola 2002 for a review). 
The organic refractory residue,
synthesized from UV photoprocessing of interstellar ice mixtures, 
provides a perfect match, better than any other hydrocarbon
analogs, to the observed 3.4$\mum$ band, 
including the 3.42$\mum$, 3.48$\mum$, and 3.51$\mum$ subfeatures 
(Greenberg et al.\ 1995).

However, the absorption spectra of organic residues display
a strong 3.0$\mum$ O-H feature and a broad 5.5--10$\mum$
band which are inconsistent with the diffuse ISM observations.
We note that these features, largely 
attributed to the combined features of 
the O-H, C=O, C-OH, C$\equiv$N, C-NH$_2$, and NH$_2$ 
stretches, bendings, and deformations, will become weaker
or even fully absent if the organics are subject to greater 
UV photoprocessing which will result in photodissociation 
and depletion of H, O, N elements. 
We also note that the organic residue samples presented 
in Greenberg et al.\ (1995) were processed at most to 
a degree resembling one cycle 
(from molecular clouds to diffuse clouds).
According to the cyclic evolutionary dust model, 
interstellar grains will undergo $\sim 50$ cycles 
before they are consumed by star formation or becomes 
a part of a comet (Greenberg \& Li 1999). 

With a rather weak 5--10$\mum$ band and absent in the 3.0$\mum$ 
O-H feature, the IR spectrum of the organic extract from 
the Murchison meteorite is also very close to the observed 
3.4$\mum$ interstellar feature (Pendleton et al.\ 1994).
We therefore adopt the Murchison meteorite 2.5--25$\mum$ 
transmission spectrum (Cronin \& Pizzarello 1990) to construct the
complex refractive index $m(\lambda)=\mre(\lambda)+i \mim(\lambda)$
for the interstellar carbonaceous organic material.

Let $\kabs(\lambda)$ be the mass absorption coefficient
at wavelength $\lambda$. For spherical grains in the Rayleigh limit 
we have $\kabs(\lambda) = 
\frac{6\pi}{\rho\lambda}~ \im\left(\frac{m^2-1}{m^2+2}\right)$
where $\rho$ is the dust mass density.
Furton, Laiho, \& Witt (1999) measured 
$\kabs(3.4\mum)\approx 1.4\times 10^3\cm^2\g^{-1}$
for a HAC sample with a hydrogen to carbon ratio of 
${\rm H/C\approx 0.5}$ considered as ``a viable analog 
to the true interstellar HAC material''.
Most recent measurements by Mennella et al.\ (2002)
found $\kabs(3.4\mum)\approx 1.6\times 10^3\cm^2\g^{-1}$
for a ${\rm H/C\approx 0.7}$ HAC sample. 
We therefore adopt $\kabs(3.4\mum) = 1.5\times 10^3\cm^2\g^{-1}$
for the interstellar carbonaceous organic dust.
Taking $\rho=1.5\g\cm^{-3}$ (Furton et al.\ 1999)
and $\mre(3.4\mum)=1.7\pm0.1$ (at 3--4$\mum$ $\mre\approx 1.81$
[Zubko et al.\ 1996]; 1.65 [Alterovitz et al.\ 1991];
1.67 [Furton et al.\ 1999]), 
we obtain $\mim(3.4\mum)\approx 0.095\pm0.007$
from $\kabs=1.4\times 10^3\cm^2\g^{-1}$.\footnote{%
  Greenberg \& Li (1996) derived $\mim(3.4\mum)\approx 0.026$
  from the mass absorption coefficient 
  $\kabs \approx 440\cm^2\g^{-1}$ measured for ``yellow stuff'' 
  (``first generation'' organic refractory
  material; Greenberg 1989b).
  }

\begin{figure}[h]
\begin{center}
\epsfig{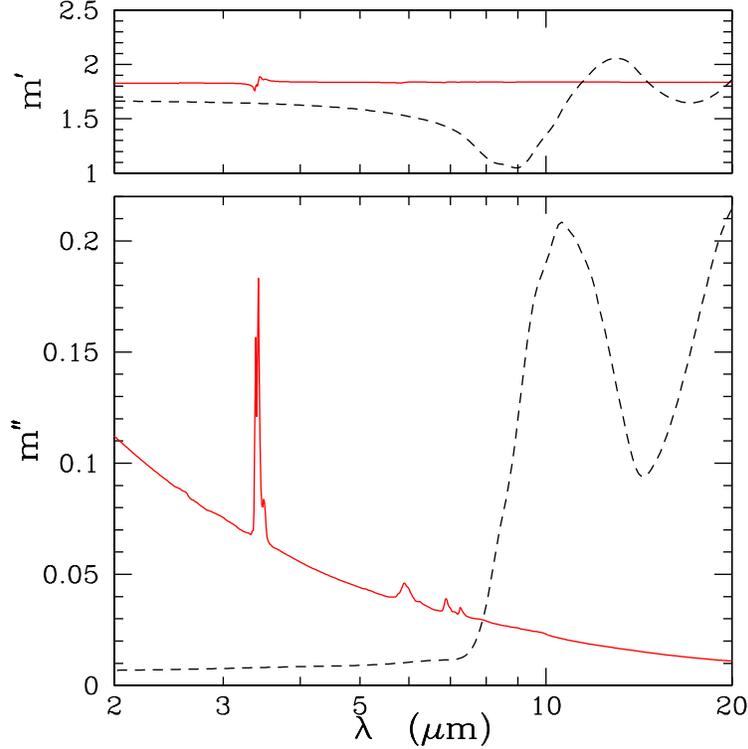}
\end{center}\vspace*{-1em}
\caption{
        \label{fig:nk}
        \footnotesize
        Refractive indices of interstellar carbonaceous 
        organic (solid line; see \S\ref{sec:opct}) 
        and silicate (dashed line; taken from Draine \& Lee [1984]) 
        dust materials dust. The imaginary part $\mim$ of 
        the ``astronomical'' silicate is reduced by a factor of 5.        
        }
\end{figure}

Neglecting reflection, the 2.5--25$\mum$ imaginary part
of the index of refraction can be calculated from the
transmission spectrum of the Murchison meteorite: 
$\mim(\lambda) = \mim(3.4\mum) 
\left(\frac{\lambda}{3.4\mum}\right)
\left(\frac{\ln\bT_\lambda}{\ln\bT_{3.4\mum}}\right)$
where $\bT_\lambda$ is the transmittance.
For $0.09\mum \simlt \lambda \simlt 1\mum$ we take
the $\mim$ values measured from the heavily processed 
organic residue (Jenniskens 1993). For $\lambda >1\mum$
we take $\mim(\lambda) = \mim(1\mum)/\lambda$ and then
smoothly add the Murchison meteorite $\mim$ to this continuum. 
Finally, the real part of the index of refraction $\mre(\lambda)$
is calculated from $\mim(\lambda)$ through the Kramers-Kronig relation.
The resulting $\mre$ and $\mim$ are shown in Figure \ref{fig:nk} 
for $2\mum < \lambda < 20\mum$.

\section{Polarization of the 3.4$\mum$ Feature\label{sec:p2a}}
To calculate the polarization of the 3.4$\mum$ C-H absorption 
feature, for simplicity, we consider spheroidal 
silicate core-carbonaceous organic mantle dust grains.  
We consider both {\it confocal} spheroidal core-mantle
grains and {\it equal-eccentricity} spheroids 
(both the core and the mantle have the same eccentricity
[i.e. the same shape]).
Undoubtedly, real interstellar grains must be more complicated.
Since our ability to compute scattering and absorption 
cross sections for nonspherical particles is extremely limited,
we use the confocal and equal-eccentricity spheroidal shapes
as a first approximation. The discrete dipole approximation
(Draine 1988) may be taken to study more complicated shapes.
Quantitative differences are expected but this will not affect
the conclusion of the paper, i.e., both the 3.4$\mum$ C-H
absorption feature and the 9.7$\mum$ Si-O feature are expected
to be polarized for nonspherical and aligned core-mantle grains. 

Let $\ac$ and $\bc$ be the {\it core} 
semiaxis along and perpendicular to the symmetry axis, respectively;
$\am$ and $\bm$ be the {\it mantle} semiaxis 
along and perpendicular to the symmetry axis, respectively;
$\ecc$ and $\ecm$ be the {\it core} and {\it mantle} eccentricities,
respectively; $\fsil$ and $\for$ ($\equiv 1-\fsil$) be the volume fractions 
of the {\it core} and the {\it mantle}, respectively.
For {\it confocal} spheroids, $\ecc$ relates to $\ecm$
through $\fsil$.

Let $\cpar$ and $\cper$ be the absorption cross sections 
for light polarized parallel and perpendicular, respectively, 
to the grain symmetry axis. For an ensemble of grains spinning
and precessing about the magnetic field, the polarization
cross section is $\cppol = \left(\cpar-\cper\right)/2$
for prolates, and $\copol = \left(\cper-\cpar\right)$ for oblates;
the absorption cross section is   
$\cabs = \left(\cpar+2\cper\right)/3 - 
\Phi \cpol \left(3-2/\cos^2\gamma\right)/6$
where $\Phi$ is the polarization reduction factor;
$\gamma$ is the angle between the magnetic field
and the plane of the sky (Lee \& Draine 1985). 
We take $\Phi=1$ and $\gamma=0$.

In the wavelength of interest here, tenth-micron 
interstellar grains are in the Rayleigh limit. 
We therefore calculate the absorption cross sections 
of confocal spheroids using the ``dipole approximation'' 
(Gilra 1972; Draine \& Lee 1984). For equal-eccentricity
spheroids we take the approach of Farafonov (2001)
and Voshchinnikov \& Mathis (1999). 
We adopt the Draine \& Lee (1984) dielectric functions 
for silicate dust.

Let $\vsil$ and $\vor$ be the silicate core and 
carbonaceous mantle volumes, respectively.
The volume fraction of the carbonaceous mantles 
$\for\, [\equiv \vor/(\vor+\vsil)]$
can be estimated from the observed 3.4$\mum$ C-H ($\tauor$)
and 9.7$\mum$ Si-O ($\tausil$) optical depths:
$\vor/\vsil \approx \left(\tauor/\tausil\right)
\left(\rho_{\rm sil}/\rho_{\rm carb}\right)\\
\left(\kappa_{\rm sil}^{9.7\mum}/\kappa_{\rm carb}^{3.4\mum}\right)
\approx 0.25$ 
where $\tauor/\tausil \approx 1/18$ (Sandford, Pendleton,
\& Allamandola 1995); $\rho_{\rm sil}$ and $\rho_{\rm carb}$
are the mass densities of the silicate ($\approx 3.5\g\cm^{-3}$)
and carbonaceous ($\approx 1.5\g\cm^{-3}$) dust;
$\kappa_{\rm sil}^{9.7\mum} (\approx 2850\cm^2\g^{-1})$ 
is the 9.7$\mum$ Si-O silicate mass absorption coefficient
(Draine \& Lee 1984);
$\kappa_{\rm carb}^{3.4\mum} (\approx 1500\cm^2\g^{-1})$ 
is the 3.4$\mum$ C-H mass absorption coefficient of 
carbonaceous organic dust (see \S\ref{sec:opct}).\footnote{%
  This appears to be consistent with the estimation of 
  Tielens et al.\ (1996) who found $\vor/\vsil \approx 0.23$.
  But we caution that much lower mass densities 
  ($\rho_{\rm sil} = 2.5\g\cm^{-3}$,
   $\rho_{\rm carb} = 1.0\g\cm^{-3}$)
   were adopted in Tielens et al.\ (1996). 
  }
However, a much thicker mantle ($\vor/\vsil \approx 1$)
is required to account for the visual/near-IR interstellar 
extinction (Li \& Greenberg 1997). 
The mass ratio of carbonaceous organics to silicates 
in the coma of comet Halley, measured {\it in situ}, 
was approximately 0.5 (Kissel \& Krueger 1987).
This also points to $\vor/\vsil \approx 1$ -- it is 
often suggested that cometary dust is of 
interstellar origin (Greenberg 1982).
In dense clouds thicker organic mantles
would be expected (but the 3.4$\mum$ C-H band strength
also changes; see \S\ref{sec:discussion}). 
Therefore, we will consider 3 cases: $\vor/\vsil = 0.25, 1, 2$. 

\begin{figure}[h] 
\begin{center}
\epsfig{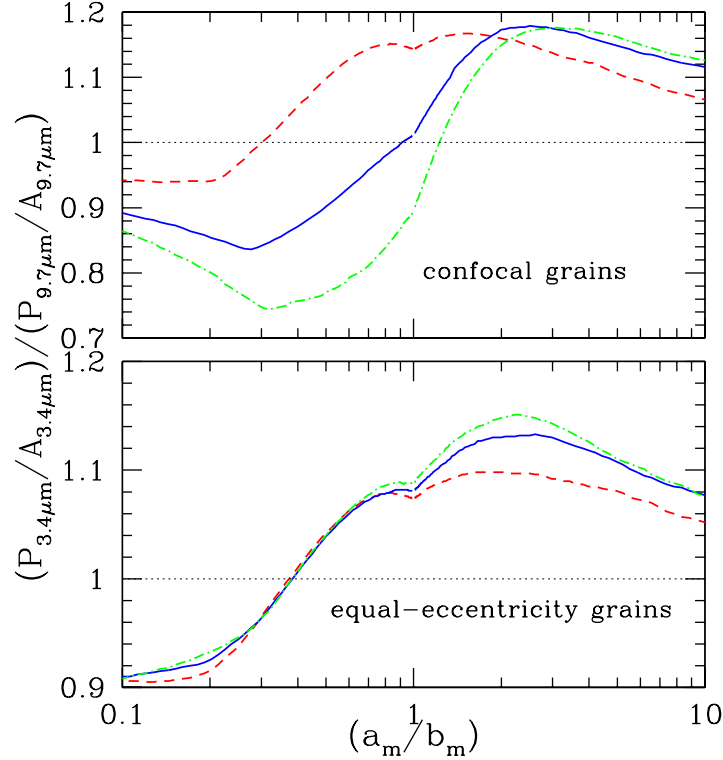}
\end{center}\vspace*{-1em}
\caption{
        \label{fig:p2a}
        \footnotesize
        Predicted ratio of the 3.4$\mum$ C-H polarization
        (relative to absorption) to the 9.7$\mum$ Si-O polarization 
        (relative to absorption) for confocal spheroids 
        (upper panel) and spheroids with an equal-eccentricity
        for both the core and the mantle (lower panel)
        as a function of mantle elongation
        for 3 different mantle thicknesses
        $\vor/\vsil = 0.25$ (dashed line), 1 (solid line), 
        and 2 (dot-dashed line). 
        }
\end{figure}

We consider spheroids with a wide range of elongations.
For a given mantle elongation $\am/\bm$,
the elongation of the core $\ac/\bc$ 
is determined from the mantle thickness$\vor/\vsil$. 
We calculate the 3.4$\mum$ C-H excess extinction
$\Ach$ and excess polarization $\Pch$, and the 9.7$\mum$ Si-O
excess extinction $\Asil$ and excess polarization $\Psil$
as a function of the mantle elongation $\am/\bm$
and mantle thickness. In Figure \ref{fig:p2a} we show
$\left(\Pch/\Ach\right)/\left(\Psil/\Asil\right)$,
the ratio of the 3.4$\mum$ polarization-to-extinction ratio 
to the 9.7$\mum$ polarization-to-extinction ratio.\footnote{%
  We note that the excess extinction and polarization are
  the peak values. The polarization peak shifts to a longer  
  wavelength relative to the extinction peak (Kobayashi et al.\ 1980).
  }

It is seen in Figure \ref{fig:p2a} that for prolate
(both confocal and equal-eccentricity) grains, the 3.4$\mum$
C-H absorption feature is polarized to a slightly higher degree  
than the 9.7$\mum$ silicate feature. For oblate grains,
the degree of polarization of the 3.4$\mum$ feature can be
lower in comparison with the 9.7$\mum$ feature, but even the
most extreme case considered in this work still has
$\left(\Pch/\Ach\right)/\left(\Psil/\Asil\right) > 75\%$.
Therefore, we can conclude that the silicate core-carbonaceous 
organic mantle model predicts a more or less similar degree of 
polarization for both the 9.7$\mum$ Si-O feature 
and the 3.4$\mum$ C-H feature.

\section{Discussion\label{sec:discussion}}
Observationally, there exists a correlation between the 9.7$\mum$ 
silicate and the 3.4$\mum$ C-H hydrocarbon optical depths 
(Sandford et al.\ 1995). Although this is consistent with 
the core-mantle model, it does not necessary establish 
a core-mantle relationship between the silicate and
hydrocarbon materials since such a correlation could
also be achieved for a proportional distribution of
these materials. In \S\ref{sec:p2a} we have seen that
both the 3.4$\mum$ feature and the 9.7$\mum$ feature are 
expected to be polarized for aligned, nonspherical silicate 
core-carbonaceous organic mantle grains. Again, a positive
detection of the 3.4$\mum$ and 9.7$\mum$ polarization
does not necessary approve the core-mantle model, 
but a positive detection of one feature together with
a negative detection of another would be a fatal judgement
for the core-mantle model.

Attempts to measure the polarization of the 3.4$\mum$ 
absorption feature ($P_{3.4\mum}^{\rm IRS7-obs}$) was recently 
made by Adamson et al.\ (1999) toward the Galactic center source 
IRS 7. They found that this feature was essentially unpolarized. 
Since {\it no} spectropolarimetric observation of the 9.7$\mum$ silicate 
absorption feature ($P_{9.7\mum}^{\rm IRS7}$) has yet been 
carried out for IRS 7, they estimated $P_{9.7\mum}^{\rm IRS7}$
from the 9.7$\mum$ silicate optical depth $A_{9.7\mum}^{\rm IRS7}$, 
{\it assuming} that the IRS 7 silicate feature is polarized to 
the same degree as the IRS 3 (another Galactic center source)
silicate feature; 
i.e., $P_{9.7\mum}^{\rm IRS7}/A_{9.7\mum}^{\rm IRS7} 
= P_{9.7\mum}^{\rm IRS3}/A_{9.7\mum}^{\rm IRS3}$
(observational data for $A_{9.7\mum}^{\rm IRS7}$, $A_{9.7\mum}^{\rm IRS3}$ 
and $P_{9.7\mum}^{\rm IRS3}$ are available).
{\it Assuming} the IRS 7 aliphatic carbon (the 3.4$\mum$ carrier) is 
aligned to the same degree as the silicate dust, they expected 
the 3.4$\mum$ polarization to be $P_{3.4\mum}^{\rm IRS7-mod}
= P_{9.7\mum}^{\rm IRS7}/A_{9.7\mum}^{\rm IRS7} 
\times A_{3.4\mum}^{\rm IRS7}$. In so doing, they found 
$P_{3.4\mum}^{\rm IRS7-obs} \ll P_{3.4\mum}^{\rm IRS7-mod}$
(Adamson et al.\ 1999). Therefore, they concluded that 
the aliphatic carbon dust is not in the form of 
a mantle on the silicate dust as suggested by 
the silicate core-carbonaceous mantle models.

We note that, to draw this conclusion, two critical 
assumptions were made:
(1)  $P_{9.7\mum}^{\rm IRS7}/A_{9.7\mum}^{\rm IRS7} 
= P_{9.7\mum}^{\rm IRS3}/A_{9.7\mum}^{\rm IRS3}$;
(2) $P_{3.4\mum}^{\rm IRS7}/A_{3.4\mum}^{\rm IRS7} 
= P_{9.7\mum}^{\rm IRS7}/A_{9.7\mum}^{\rm IRS7}$.
Although the 2nd assumption does not seem to significantly
deviate from the detailed calculations 
[$0.7 \simlt  \left(P_{3.4\mum}/A_{3.4\mum}\right) 
/\left(P_{9.7\mum}/A_{9.7\mum}\right)\simlt 1.2$
for reasonable dust parameters; 
see \S\ref{sec:p2a} and Figure \ref{fig:p2a}],
the 1st assumption is questionable and needs observational support.
We urgently need spectropolarimetric observations of IRS 7
in the 9.7$\mum$ region or of IRS 3 in the 3.4$\mum$ region. 
If we take the most likely dust parameters
(1) $\am/\bm=1/2$ oblate\footnote{%
  The $a/b=1/2$ oblate shape was shown to provide a better match 
  than any other shapes to the 3.1$\mum$ ice polarization 
  feature (Lee \& Draine 1985) and the 9.7$\mum$ silicate 
  polarization feature (Hildebrand \& Dragovan 1995) toward
  the Becklin-Neugebauer object. 
  }
and (2) $\vor/\vsil=1$ (Li \& Greenberg 1997), our model 
calculations give $\left(P_{3.4\mum}/A_{3.4\mum}\right) 
/\left(P_{9.7\mum}/A_{9.7\mum}\right) \simgt 0.9$ 
(see Figure \ref{fig:p2a}),
we therefore expect an excess polarization $< 0.8\%$
for the IRS 7 9.7$\mum$ silicate feature,
and $> 0.23\%$ for the IRS 3 3.4$\mum$ C-H feature.

It is worth noting that the absorption features
of ices -- the ``precursor'' of organic residue
-- are seen in polarization in various sources 
(see Li \& Greenberg 2002 for a recent review). 
The ice polarization feature toward 
the  Becklin-Neugebauer object was well fitted by ice-coated 
grains (Lee \& Draine 1985), suggesting a core-mantle 
grain morphology. 

The exact nature of the 3.4$\mum$ feature carrier 
remains a subject of debate (see Pendleton \& Allamandola 
2002, Li \& Greenberg 2002). The carbonaceous organic 
refractory proposal recently receives further support
from the very recent discovery of a 6.0$\mum$ feature 
in dense clouds which was attributed to organic refractory
by Gibb \& Whittet (2002). 
They found that its strength is correlated with the 
4.62$\mum$ OCN$^{-}$ (``XCN''; Schutte \& Greenberg 1997) 
feature which is considered to be a diagnostic of energetic 
processing. It would be interesting to see whether heavily
processed organic residue is able to provide a close match 
to the 5.85$\mum$ and 7.25$\mum$ C-H deformation bending modes
seen toward the Galactic center source Sgr A$^{\ast}$ as well
(Chiar et al.\ 2000).

The 3.4$\mum$ feature consists of three subfeatures at 
${\rm 2955\,cm^{-1}}$ (3.385$\mu$m), 
${\rm 2925\,cm^{-1}}$ (3.420$\mum$), 
and ${\rm 2870\,cm^{-1}}$ (3.485$\mum$)
corresponding to the symmetric and asymmetric C-H stretches 
in CH$_3$ and CH$_2$ groups in aliphatic hydrocarbons which 
must be interacting with other chemical groups.
The amount of carbonaceous material responsible for the  
3.4$\mum$ feature is strongly dependent on the nature 
of the chemical groups attached to the aliphatic carbons. 
For example, each carbonyl (C=O) group reduces its corresponding 
C-H stretch strength by a factor of $\sim 10$ (Wexler 1967). 
Furthermore, not every carbon is attached to a hydrogen as in 
saturated compounds; instead, some form an aromatic structure
(see Pendleton \& Allamandola 2002).
The fact that the 3.4$\mum$ absorption is not observed 
in molecular cloud may possibly be attributed to 
dehydrogenation or oxidation (formation of carbonyl) of the 
organic refractory mantle by accretion and photoprocessing in 
the dense regions (see Greenberg \& Li 1999)
-- the former reducing the absolute number of CH stretches, 
the latter reducing the CH stretch strength by a factor of 10 
(Wexler 1967) -- the 3.4$\mum$ feature would be reduced 
{\it per unit mass} in molecular clouds. 

At this moment, we are not at a position to rule out other 
dust sources as the interstellar 3.4$\mum$ feature carrier.
This feature has also been detected in a carbon-rich 
protoplanetary nebula CRL 618 (Lequeux \& Jourdain de Muizon 1990; 
Chiar et al.\ 1998) with close resemblance to the interstellar 
feature. However, after ejection into interstellar space, 
the survival of this dust in the diffuse ISM is questionable 
(see Draine 1990). 
 
\section{Summary\label{sec:sum}}
The polarization of the 3.4$\mum$ aliphatic hydrocarbon
absorption feature is modelled in terms of the 
silicate core-organic hydrocarbon mantle interstellar dust 
model. For aligned, nonspherical core-mantle grains, 
both the 3.4$\mum$ C-H feature and the 9.7$\mum$ silicate
feature are expected to be polarized, although unlikely 
to the same degree. The non-detection of the 3.4$\mum$ 
polarization in the Galactic center source IRS 7 itself
is not sufficient enough to rule out the core-mantle model 
due to the lack of spectropolarimetric observation of 
this source in the 9.7$\mum$ region.   
We call for spectropolarimetric observations of the IRS 7
9.7$\mum$ band and the IRS 3 3.4$\mum$ band. This will
promise a powerful test on the core-mantle dust model.

\acknowledgments
A. Li (the first author of this paper) were deeply saddened by 
the passing away of Prof. J. Mayo Greenberg (the second author) 
on November 29, 2001. As a pioneer in the fields of 
cosmic dust, comets, astrochemistry, astrobiology 
and light scattering, Mayo will be remembered forever. 
We thank Prof. B.T. Draine for his invaluable comments and suggestions;
Prof. J.I. Lunine for helpful discussions;
and Dr. R.H. Lupton for the availability of the SM plotting package. 
A. Li thanks The University of Arizona for the Arizona Prize 
Postdoctoral Fellowship in Theoretical Astrophysics.
This research was supported in part by
NASA grant NAG5-10811 and NSF grant AST-9988126.

\end{document}